\documentclass{ws-ijbc}
\usepackage{ws-rotating}     
\usepackage{graphicx}
\usepackage{epstopdf}
\usepackage{booktabs}
\usepackage{amssymb}
\usepackage{epsfig}
\usepackage{amsmath}
\usepackage{float}
\usepackage{subfigure}
\usepackage{booktabs}
\usepackage{colortbl}
\usepackage{color}
\usepackage[utf8]{inputenc}
\usepackage[english]{babel}

\begin{document}

\catchline{}{}{}{}{} 

\markboth{Author's Name}{Paper Title}

\title{Computing complex horseshoes by means of piecewise maps}

\author{Álvaro G. López, \'{A}lvar Daza, Jes\'{u}s M. Seoane}\address{Nonlinear Dynamics,
Chaos and Complex Systems Group, Departamento de F\'{i}sica, \\
Universidad Rey Juan Carlos, Tulip\'{a}n s/n, 28933 M\'{o}stoles,
Madrid, Spain \\ alvaro.lopez@urjc.es}

\author{Miguel A. F. Sanjuán } \address{Nonlinear Dynamics,
Chaos and Complex Systems Group, Departamento de F\'{i}sica, \\
Universidad Rey Juan Carlos, Tulip\'{a}n s/n, 28933 M\'{o}stoles,
Madrid, Spain}
\address{Department of Applied Informatics, Kaunas 
University of Technology, \\ Studentu 50-415, Kaunas LT-51368, Lithuania}

\maketitle

\begin{history}
\received{(to be inserted by publisher)}
\end{history}

\begin{abstract}
A systematic procedure to numerically compute a horseshoe map is presented. This new method uses piecewise functions and expresses the required operations by means of elementary transformations, such as translations, scalings, projections and rotations. By repeatedly combining such transformations, arbitrarily complex folding structures can be created. We show the potential of these horseshoe piecewise maps to illustrate several central concepts of nonlinear dynamical systems, as for example the Wada property.
\end{abstract}


\section{Introduction} \label{sec:intro}

The Smale horseshoe map is one of the hallmarks of chaos. It was devised in the 1960's by Stephen Smale \cite{smale} to reproduce the dynamics of a chaotic flow in the neighborhood of a given periodic orbit. It describes in the simplest way the dynamics of homoclinic tangles, which were encountered by Henri Poincar\'{e} \cite{poinca} and were intensively studied by George Birkhoff \cite{birkho}, and later on by Mary Catwright and John Littlewood, among others \cite{carlit1,carlit2,levin}. 

\begin{figure}
\centering
\bigskip
\bigskip
   \includegraphics[width=0.65\linewidth] {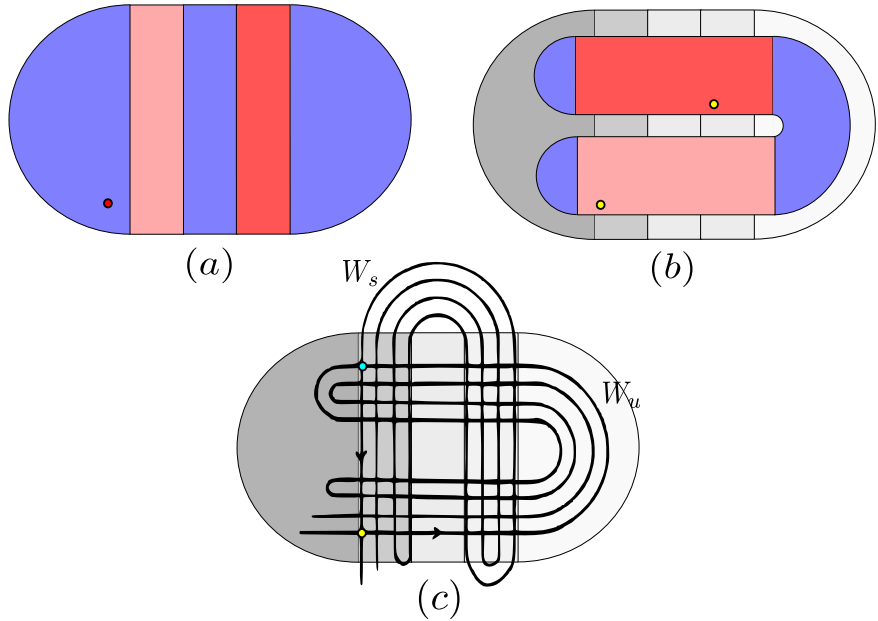} 

\caption{(a) The original set $S$ with the shape of a stadium whose iterate by means of the Smale horseshoe map $H$ is embedded into itself. In red we show an attracting fixed point. (b) The horseshoe as the first iteration of the stadium $H(S)$. The colors match the initial set and the images through the map. Two saddle fixed points are shown in yellow. (c) The invariant manifolds associated with one of the saddles shown before. In blue, a transverse homoclinic intersection between the stable and the unstable manifolds.}
\label{fig:sh0}
\end{figure}

The horseshoe map is commonly represented starting with a set with the shape of a stadium. This set is flattened along its shortest side and stretched along the orthogonal direction, which corresponds to the largest side of the stadium. Finally, the resulting set is folded acquiring the shape of a horseshoe and embedded in the original set. A two-dimensional fixed point theorem demonstrates that every continuous map defined on the plane, which iterates some set transversally to itself, contains a fixed point \cite{york}. Since the horseshoe has two legs, there must be at least two fixed points in the original domain (see Figs.~\ref{fig:sh0}(a) and (b)). These two fixed points are saddles, whose invariant manifolds take after the iterations of the horseshoe, as shown in Fig.~\ref{fig:sh0}(c). A third fixed point outside the region defined by the three vertical strips represents an attractor, as depicted in red color in Fig.~\ref{fig:sh0}(a). All the initial conditions belonging to the stadium, except for an invariant set of zero Lesbesgue measure, end up in such attractor.

The dynamics of the horseshoe map has been widely studied using Markov partitions and symbolic dynamics \cite{adler,devaney}. Also simpler discontinuous maps, as for example the baker map \cite{halmos}, can be devised, which  allow to easily compute their Lyapunov exponents in order to prove the existence of chaotic trajectories \cite{york}. Nevertheless, as far as we are concerned, a set of equations that permit to compute a continuous map that represents a horseshoe with any desired number of foldings is still missing. The relevance of such a mapping lies in the fact that it can be used to study properties of chaotic dynamical systems and, perhaps more importantly nowadays, to develop or simplify control methods based on the dynamics of the horseshoe \cite{partcon}.

In the present work we introduce a mechanistic procedure to construct two-dimensional continuous mappings that allow to numerically compute horseshoes with any desired number of foldings. The paper is organized as follows. In Sec.~\ref{sec:fsf} we describe four two-dimensional continuous maps, which can be composed to obtain the Smale horseshoe map. In Sec.~\ref{sec:exa} we present an example that shows how to extend the map to represent more complicated horseshoes, with more than one folding. Finally, in Sec.~\ref{sec:pro} we present several applications of these new piecewise mappings.  
\begin{figure}
\centering
\bigskip
\bigskip
   \includegraphics[width=0.75\linewidth] {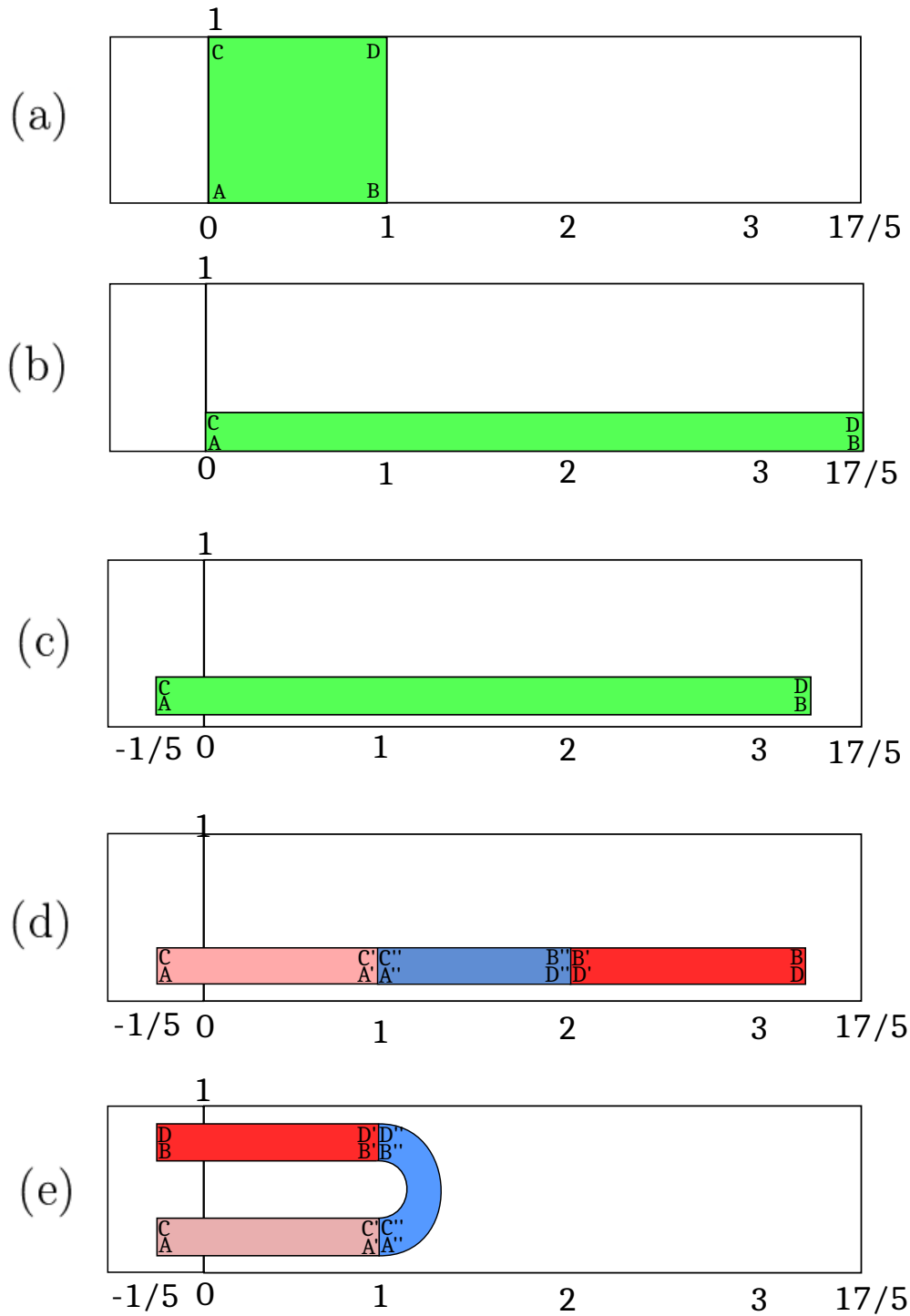} 

\caption{(a) The unit square $Q\subset \mathbb{R}^{2}$. (b) The result of contracting and stretching the unit square. (c) Translation of the set obtained in (b). (d)-(e) The unit square is folded, obtaining the shape of a horseshoe.}
\label{fig:hor1}
\end{figure}

\section{Flatten, stretch and fold}\label{sec:fsf}

A hyperbolic chaotic flow consists in an unstable and bounded motion, for which it is necessary that some directions in phase space are locally expansive (unstable) and some others are contractive (stable). In particular, if an ensemble of initial conditions in phase space is considered, after a very short time, this set will have expanded along some directions and contracted along some others. For simplicity, suppose that the dynamical system is three-dimensional and that a Poincar\'e map can be constructed at the plane $z=0$. As shown in Fig.~\ref{fig:hor1}(a), we begin with a square $Q \subset \mathbb{R}^{2}$ as the initial set. Then, suppose that a forward iteration of the map expands this set linearly and homogeneously in the direction of the $x$-axis by some amount and contracts it in the other direction $y$ in a similar way. Obviously, we obtain as a result a rectangle, as depicted in Fig~\ref{fig:hor1}(b).

If we carry out this linear transformation repeatedly, we would asymptotically obtain an unbounded set in the direction of the $x$-axis, isomorphic to the real line. Therefore, if we want to keep the evolving set bounded, we need to provide a mechanism to bend the set against the expanding direction. It is precisely the nonlinearity that introduces the curvature required for the bending, maintaining the dynamics trapped in some region of phase space. This requirement is essential, since smooth linear transformations transform parallel lines into parallel lines. In summary, a continuous chaotic dynamical system requires at least three ingredients: a contraction and a stretch (saddles), which can be carried out by simple linear transformations, and a folding, for which a nonlinear transformation is necessary. 
 
Based on these ideas, we can build a horseshoe in a very simple manner by a composition of elementary transformations. We follow the steps represented in Fig.~\ref{fig:hor1}. First of all, we start with the unit square $Q=[0,1] \times [0,1]$ and carry out the flattening and the stretching shown in Fig.~\ref{fig:hor1}(b), by means of the function $H_{1}(x,y)$, which is defined as follows
\begin{equation}
H_{1}(x,y)=\left\{ \begin{array}{rcl}
(x,y/4) & \mbox{for} & x<0 \\ 
(17/5 x,y/4) & \mbox{for} & 0 \leq x\leq 1 \\
(x,y/4)+(12/5,0) & \mbox{for} & x>1
\end{array}\right.\label{eq:1}
\end{equation}

Note that the stretching only applies in the region $0 \leq x\leq 1$ and that we have added a translation for $x>1$. The reason for such restriction is that we want to keep the shape of the blue folding, which lies out of the unit square after the first iteration (see Fig.~\ref{fig:hor1}(e)), when we apply the second iteration of the map. In this manner, we avoid its expansion along the $x$-axis. Instead of stretching it, we simply translate it in such a way that, when computing the second iteration of the map, it is placed at the end of the flattened and stretched green rectangle appearing in Fig.~\ref{fig:hor1}(c).
\begin{figure}
\centering
\bigskip
\bigskip
   \includegraphics[width=0.65\linewidth] {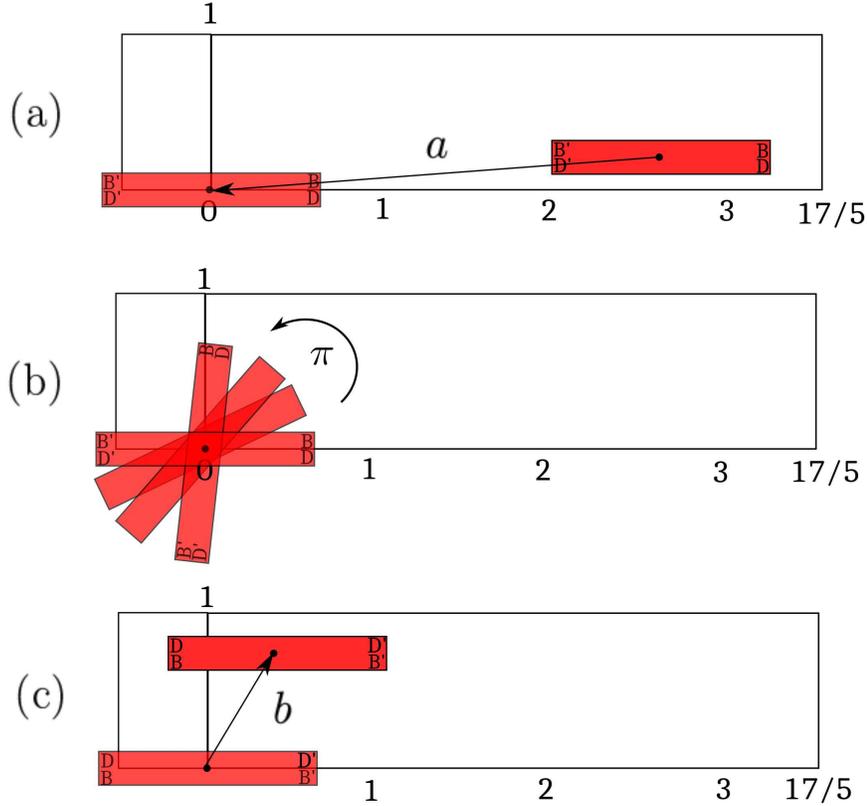} 

\caption{(a) First step associated to the folding of the red region in Fig.~\ref{fig:hor1}(e), consisting in a translation. (b) A rotation of $\pi$ radians. (c) A translation placing the red rectangle symmetrical to the other clearer red rectangle with respect the axis $y=1/2$.}
\label{fig:hor2}
\end{figure}

Secondly, we perform a translation over the whole set, as depicted in Fig.~\ref{fig:hor1}(c), lifting up the whole thin rectangle and moving it to the left, so that it lies slightly above the $x$-axis and runs through the $y$-axis. This affine transformation can be written as
\begin{equation}
H_{2}(x,y)=(x,y)+(-1/5,1/8).
\end{equation}

\begin{figure}
\centering
\bigskip
\bigskip
   \includegraphics[width=0.75\linewidth] {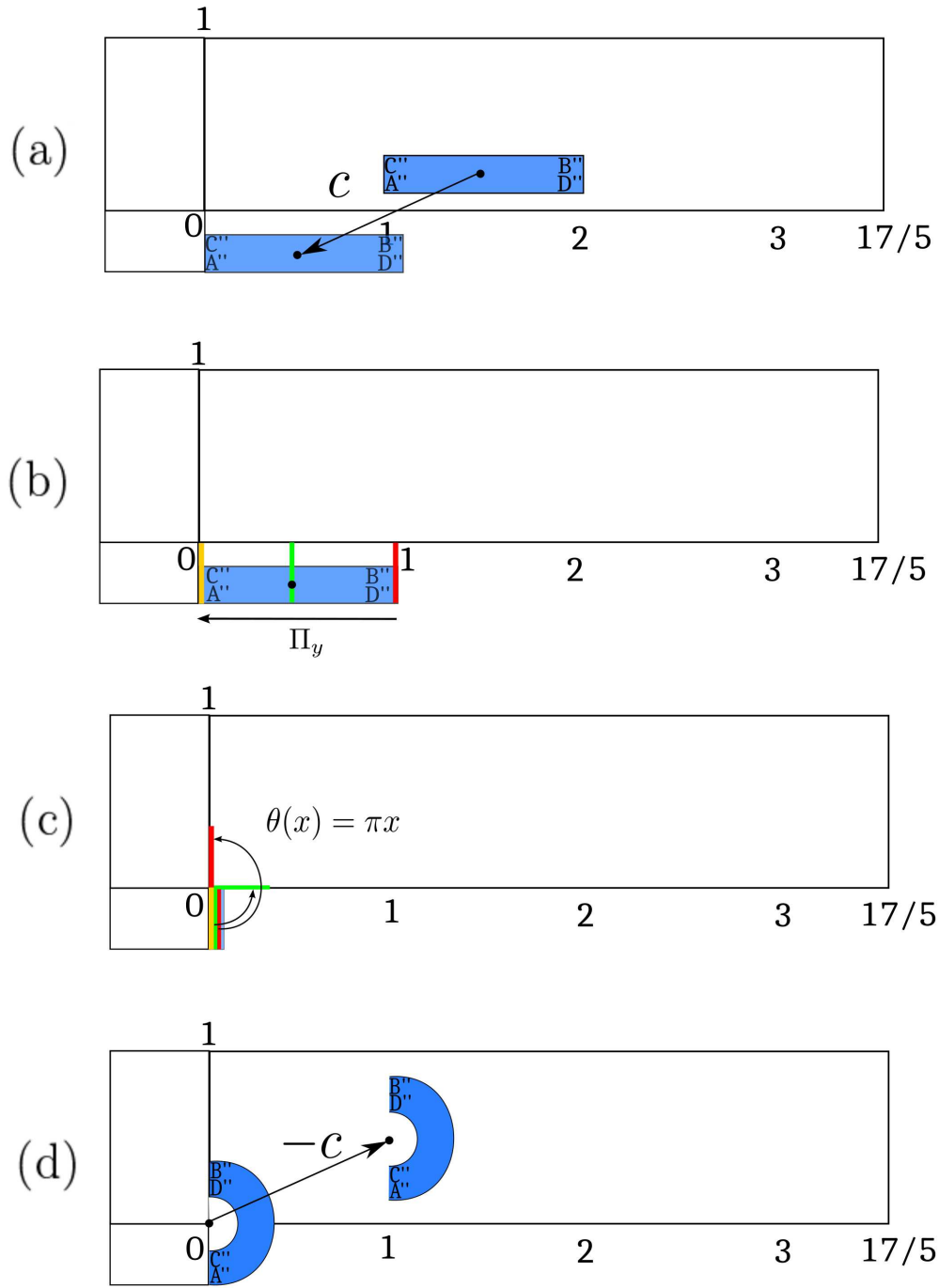} 

\caption{(a) A translation of the blue region in Fig.~\ref{fig:hor1}(e), required to achieve the folding. (b) A projection of the translated blue rectangle. (c) A nonlinear rotation generating the curvature of the folding. (d) A translation to place the folded region in its right place.}
\label{fig:hor3}
\end{figure}

Now it comes the most difficult part, which is the folding shown in Figs.~\ref{fig:hor1}(d) and (e). We perform it separately in two steps, one related to the red set, and another for the blue one. Firstly, we apply a translation $T(a)$ to the red rectangle appearing in Fig.~\ref{fig:hor1}(d) to place it at the origin, where $a$ represents the vector along which the translation is carried out. Then, we apply the counterclockwise rotation $R(\theta)$ by an angle $\theta=\pi$ radians to the rectangle. Finally, we perform one more translation $T(b)$ of the rectangle by means of the vector $b$, so that it lies symmetrical to the other light red rectangle (see Fig.~\ref{fig:hor1}(e)) with respect to the line $y=1/2$. These operations can be written as
\begin{equation}
T(b)R(\theta)T(a)x=\begin{pmatrix}
\cos\theta & -\sin\theta\\
\sin\theta & \cos\theta 
\end{pmatrix}
\begin{pmatrix}
x+a_{x}\\
y+a_{y} 
\end{pmatrix}
+
\begin{pmatrix}
b_{x}\\
b_{y} 
\end{pmatrix},
\end{equation}
where the vector parameters $a=(-13/5,-1/4)$ and $b=(2/5,3/4)$ have been chosen, in accordance with previous parameter values appearing in the transformation $H_{1}$. The whole transformation is shown in Fig.~\ref{fig:hor2}. Then, we proceed with the blue region, which is more complicated than the red, since it corresponds to the nonlinear folding. We start by applying a translation $T(c)$ to the blue rectangle shown in Fig.~\ref{fig:hor3}(a) along the direction defined by the vector $c$. Then, we project the rectangle on the $y$-axis by means of the operator $\Pi_{y}$, which corresponds to a projector defined as
\begin{equation}
\Pi_{y}=\begin{pmatrix}
& 0  &&  0 &\\
& 0  &&  1 & 
\end{pmatrix}.
\end{equation}

Such projection is shown in Fig.~\ref{fig:hor3}(b). Finally, we have to perform a rotation $R(\theta)$ of an angle $\theta$ depending on the position on the $x$-axis, so that the thin vertical yellow strip appearing in Fig.~\ref{fig:hor3}(b) and Fig.~\ref{fig:hor3}(c) does not rotate, the green vertical strip rotates $\pi/2$ radians and the red one rotates $\pi$ radians. This can be achieved by a linear dependence between the angle and the $x$ coordinate in the form $\theta(x)=\pi x$. With this linear dependence of the angle with the coordinate $x$, the rotation $R(\theta(x))$ becomes nonlinear. Finally, and as shown in Fig.~\ref{fig:hor2}(d), we perform the translation $T^{-1}(c)$ to place the folded part back in its original place. In summary, the whole transformation is $T^{-1}(c)R(\theta(T(c)x))\Pi_{y}T(c)x$ and, in matrix form, it explicitly reads
\begin{equation}
\begin{pmatrix}
\cos\pi(x+c_{x}) & -\sin\pi(x+c_{x})\\
\sin\pi(x+c_{x}) & \cos\pi(x+c_{x}) 
\end{pmatrix}
\begin{pmatrix}
0 & 0\\
0 & 1 
\end{pmatrix}
\begin{pmatrix}
x+c_{x}\\
y+c_{y} 
\end{pmatrix}
-
\begin{pmatrix}
c_{x}\\
c_{y} 
\end{pmatrix},
\end{equation}
where $c=(-1,-1/2)$, as shown in Fig.~\ref{fig:hor2}(a). The map that performs all these algebraic operations can be written as
\begin{equation}
H_{3}(x,y)=\left\{ \begin{array}{rcl}
(x,y) & \mbox{for} & x\leq 1 \\ 
(y-1/2)(-\sin(\pi(x-1)),\cos(\pi (x-1)))+(1,1/2) & \mbox{for} & 1 < x< 2 \\
-((x,y)-(13/5,1/4))+(2/5,3/4) & \mbox{for} & x \geq 2
\end{array}\right. .
\label{eq:6}
\end{equation}

Finally, for aesthetic purposes, to keep the shape of the regions in the negative part of the $x$-axis unaltered, we perform a contraction of the same size that the one performed in the first contraction in the $y$-axis, but now for negative values of the $x$-axis. Therefore, this fourth transformation reads
\begin{equation}
H_{4}(x,y)=\left\{ \begin{array}{rcl}
(x/4,y) & \mbox{for} & x< 0 \\ 
(x,y) & \mbox{for} & x \geq 0
\end{array}\right. .
\end{equation}

\begin{figure}
\centering
\begin{tabular}{cc}

\subfigure[]{
   \includegraphics[width=0.45\linewidth] {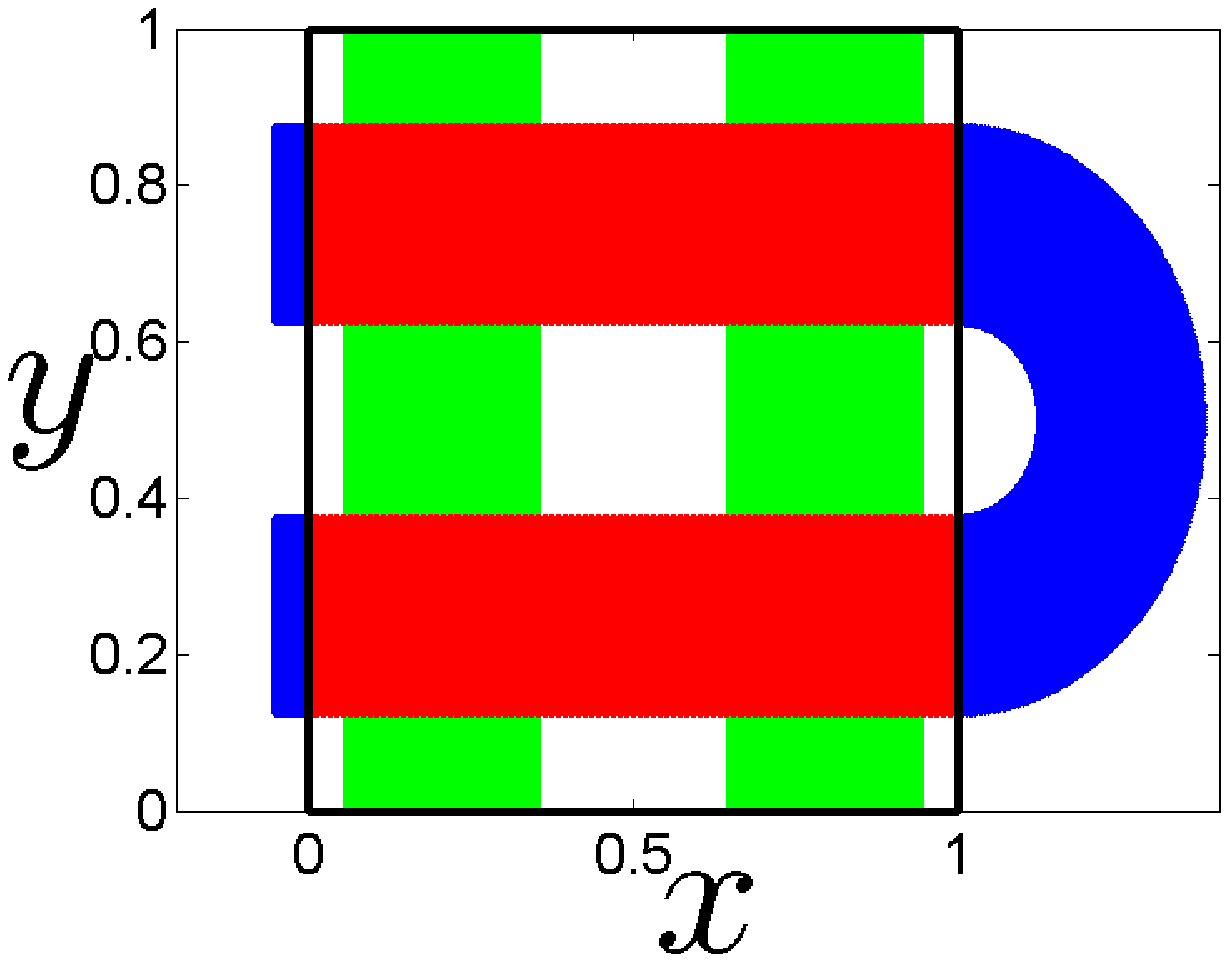}
   \label{subfig:h1}
   } &
\subfigure[]{
   \includegraphics[width=0.45\linewidth] {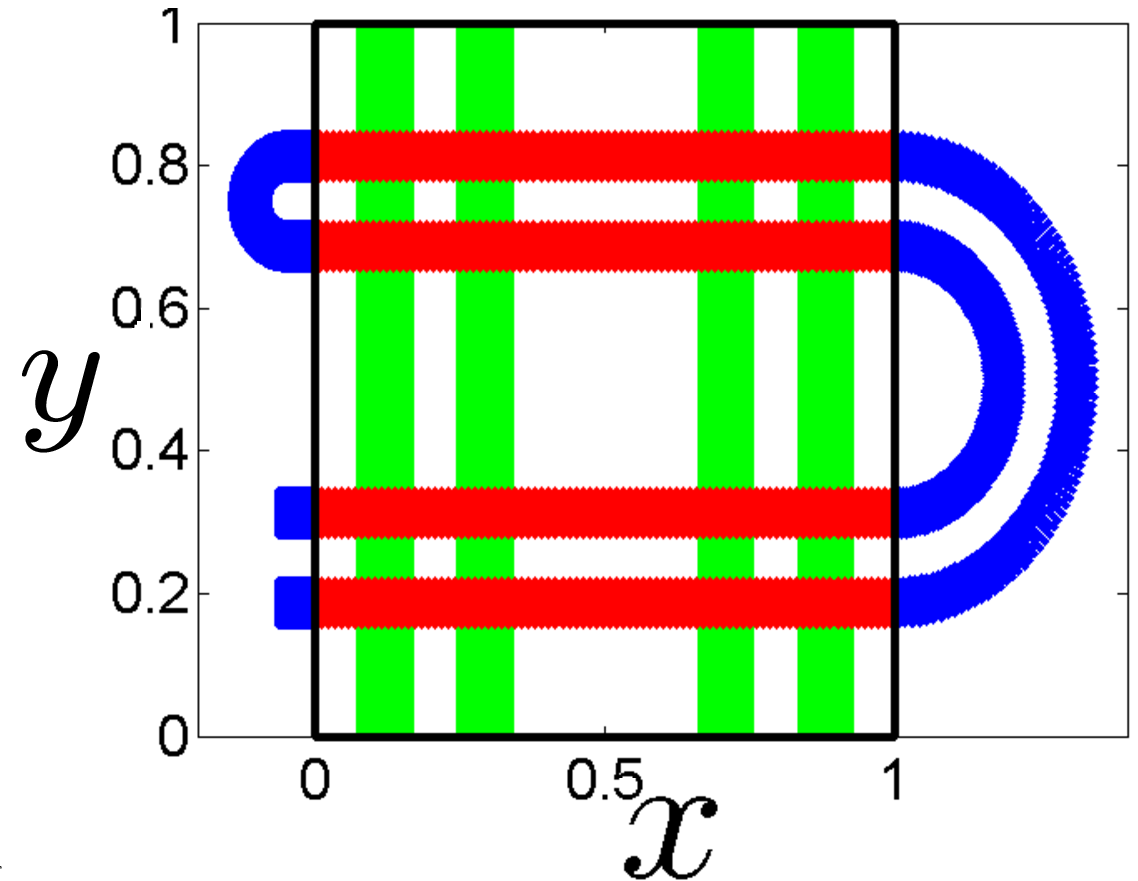}
   \label{subfig:h2}
   } \\
\subfigure[]{
   \includegraphics[width=0.49\linewidth] {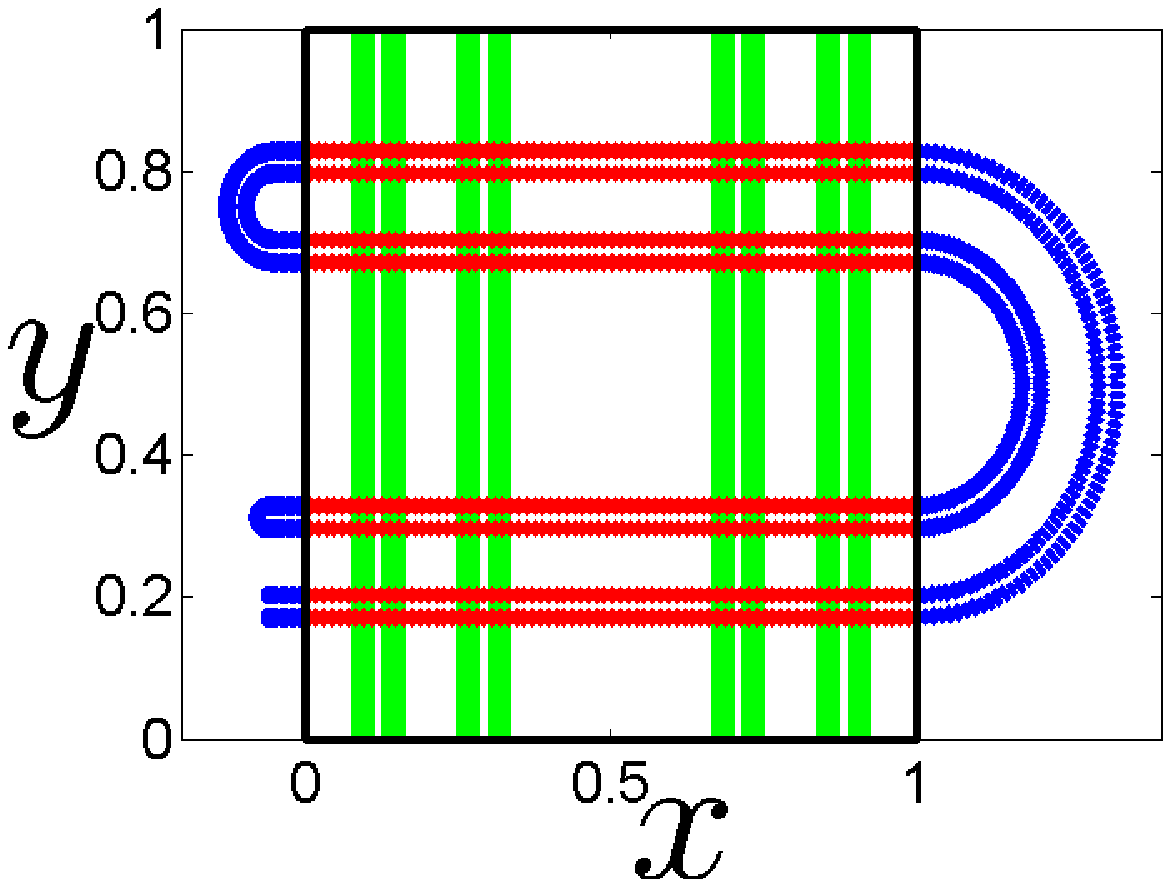}
   \label{subfig:h3}
   } &
\subfigure[]{
   \includegraphics[width=0.45\linewidth] {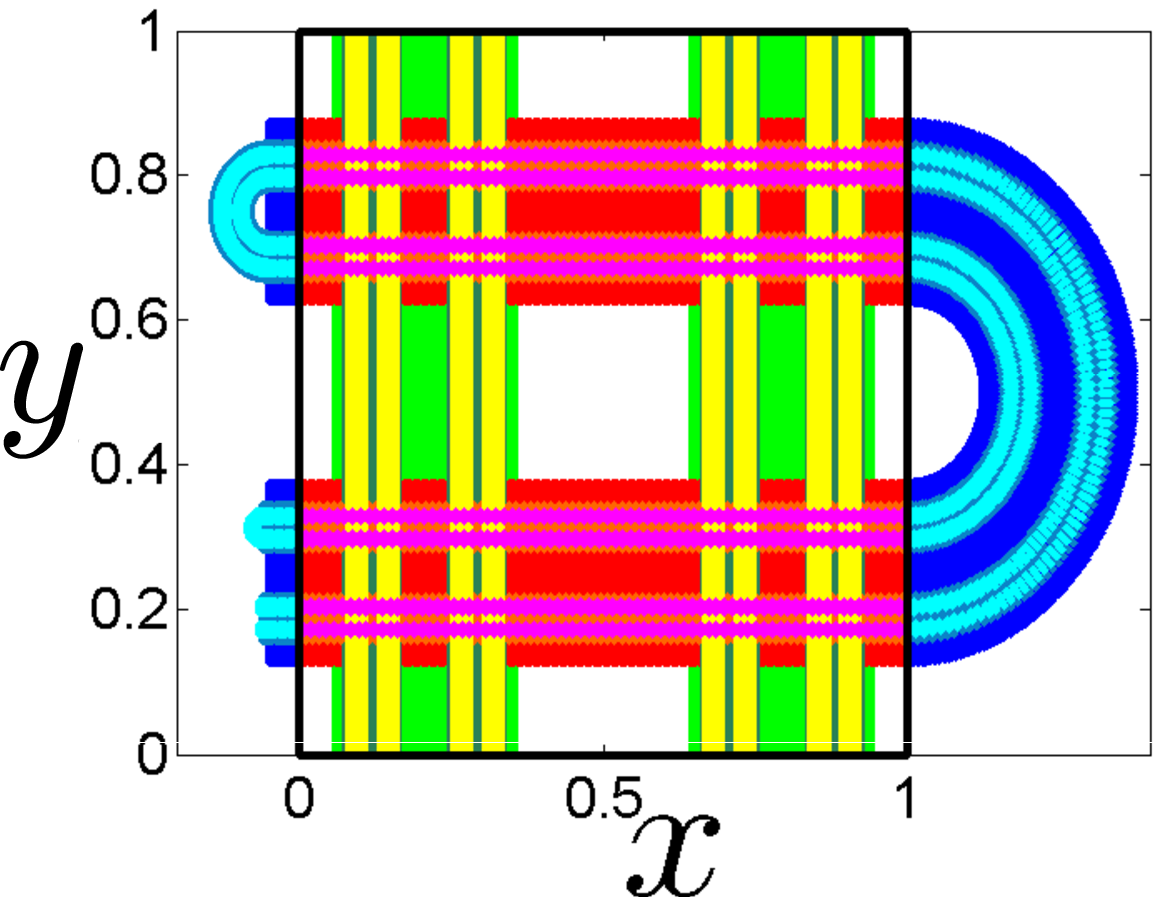}
   \label{subfig:h4}
   }

\end{tabular}
\caption{(a) The first iteration of the unit square under the piecewise horseshoe map $H(Q)$ is shown in red and blue. (b) The second iteration $H^{2}(Q)$. (c) The third iteration $H^{3}(Q)$. (d) A superposition of (a), (b) and (c). The green vertical strips in (a), (b) and (c) represent the preimages of the red rectangles, mathematically expressed as $H^{-n}(H^{n}(Q) \cap Q)$, for $n=1,2,3$, respectively.}
\label{fig:smho}
\end{figure}

The horseshoe map is simply the composition of these four transformations $H=H_{4}\circ H_{3}\circ H_{2}\circ H_{1}$. We recall that this map $H$ is continuous and one-to-one, but not differentiable, since $H_{3}$ is not differentiable at the points $(1,y)$ and $(2,y)$, which are the points delimiting the folded rectangle. This lack of smoothness contrasts with the original horseshoe conceived by Stephen Smale \cite{smale}. In Fig.~\ref{fig:smho} we take the unit square and iterate it under the map $H$, showing its successive iterations. Even though the equations of the map (Eq.~\eqref{eq:1} and Eq.~\eqref{eq:6}) look cumbersome, their interpretation in terms of elementary linear transformations and their numerical programming are rather simple. This fact has the advantage of allowing us to develop more sophisticated horseshoes in the following sections.

As we have designed it, the map contracts the area of the original unit square, since the contraction defined in $H_{1}$ is more pronounced than the stretching appearing in the same transformation. Moreover, the folding appearing in Fig.~\ref{fig:hor3} also contracts area. Nevertheless, the design of a similar map preserving the area of the unit square can be done by redefining the relation between the stretching and the contraction appearing in Eq.~\eqref{eq:1}. The map exhibits transient chaos. All the trajectories except for the chaotic saddle, a set of zero Lesbesgue measure, end in a stable fixed point $x^{*}=(-3/20,5/20)$ placed out of the unit square $Q=[0,1] \times [0,1]$. Such invariant set is computed in Sec.~\ref{sec:pro}.

\section{Horseshoes with more than one folding} \label{sec:exa}

Following the same recipe, we can construct more sophisticated schemes with an arbitrary number of foldings. For instance, a double horseshoe has been proposed to explain the fractal basin boundaries separating two attractors \cite{macdonal,aguirre}. This situation, also associated with transient chaotic dynamics, can be reproduced following the same procedure used in the previous section. A possible map capable of representing fractal basin boundaries results from the composition of the four following piecewise maps
\begin{equation}
G_{1}(x,y)=\left\{ \begin{array}{rcl}
(x,y/5) & \mbox{for} & x<0 \\ 
(27/5 x,y/5) & \mbox{for} & 0 \leq x\leq 1 \\
(x,y/5)+(22/5,0) & \mbox{for} & x>1
\end{array}\right. ,
\end{equation}  
\begin{equation}
G_{2}(x,y)=(x,y)+(-1/5,1/10),
\end{equation}
\begin{equation}
G_{3}(x,y)=\left\{ \begin{array}{rcl}
(x,y) & \mbox{for} & x\leq 1 \\ 
(y-7/20)(-\sin(\pi(x-1)),\cos(\pi (x-1)))+(1,7/20) & \mbox{for} & 1 < x \leq 2 \\
-((x,y)-(5/2,1/5))+(1/2,5/10) & \mbox{for} & 2 < x \leq 3 \\
(y-7/20)(\sin(\pi(x-3)),\cos(\pi (x-3)))+(0,13/20) & \mbox{for} & 3 < x \leq 4 \\
(x-4,y-6/10) & \mbox{for} & x > 4
\end{array}\right. ,
\end{equation}
\begin{equation}
G_{4}(x,y)=\left\{ \begin{array}{rcl}
(x/5,y) & \mbox{for} & x < 0 ~ y<7/20  \\ 
((x-1)/5+1,y) & \mbox{for} & x > 1 ~ y>13/20 \\
(x,y) & \mbox{else}
\end{array}\right. .
\end{equation}

\begin{figure}
\centering
\begin{tabular}{cc}

\subfigure[]{
   \includegraphics[width=0.475\linewidth] {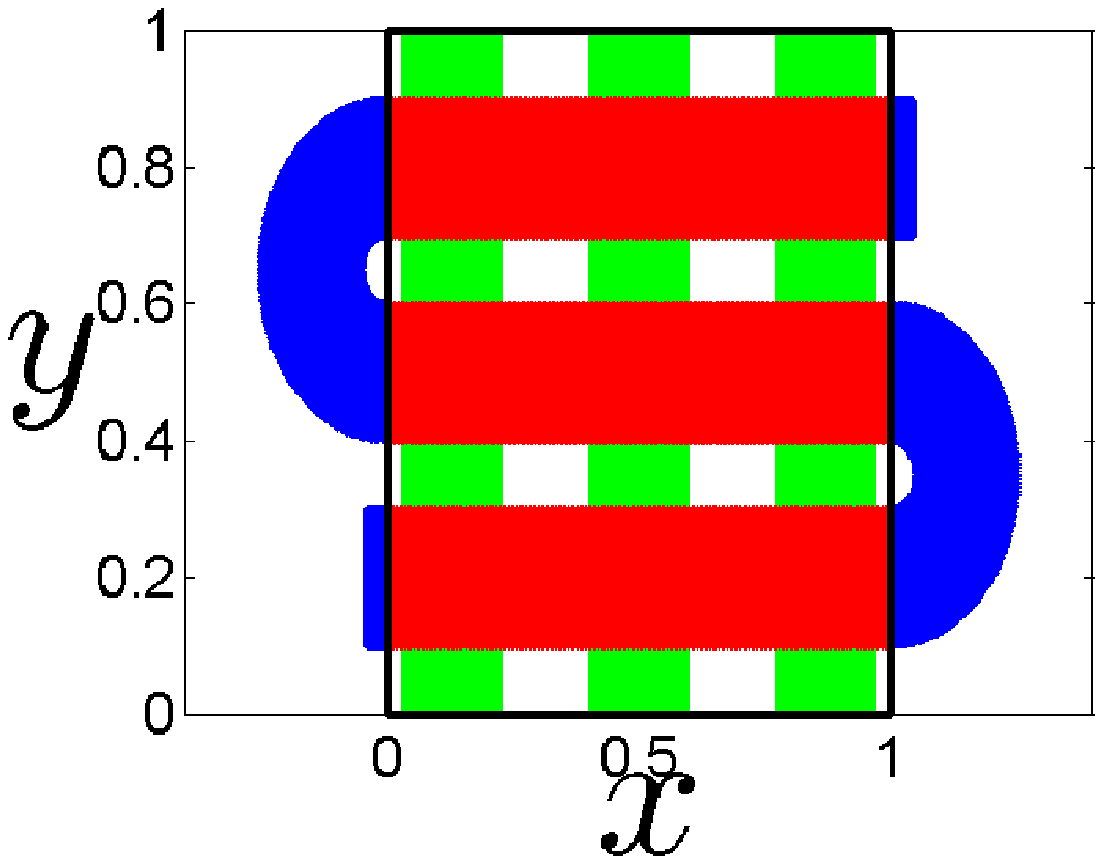}
   \label{subfig:h21}
   } &
\subfigure[]{
   \includegraphics[width=0.45\linewidth] {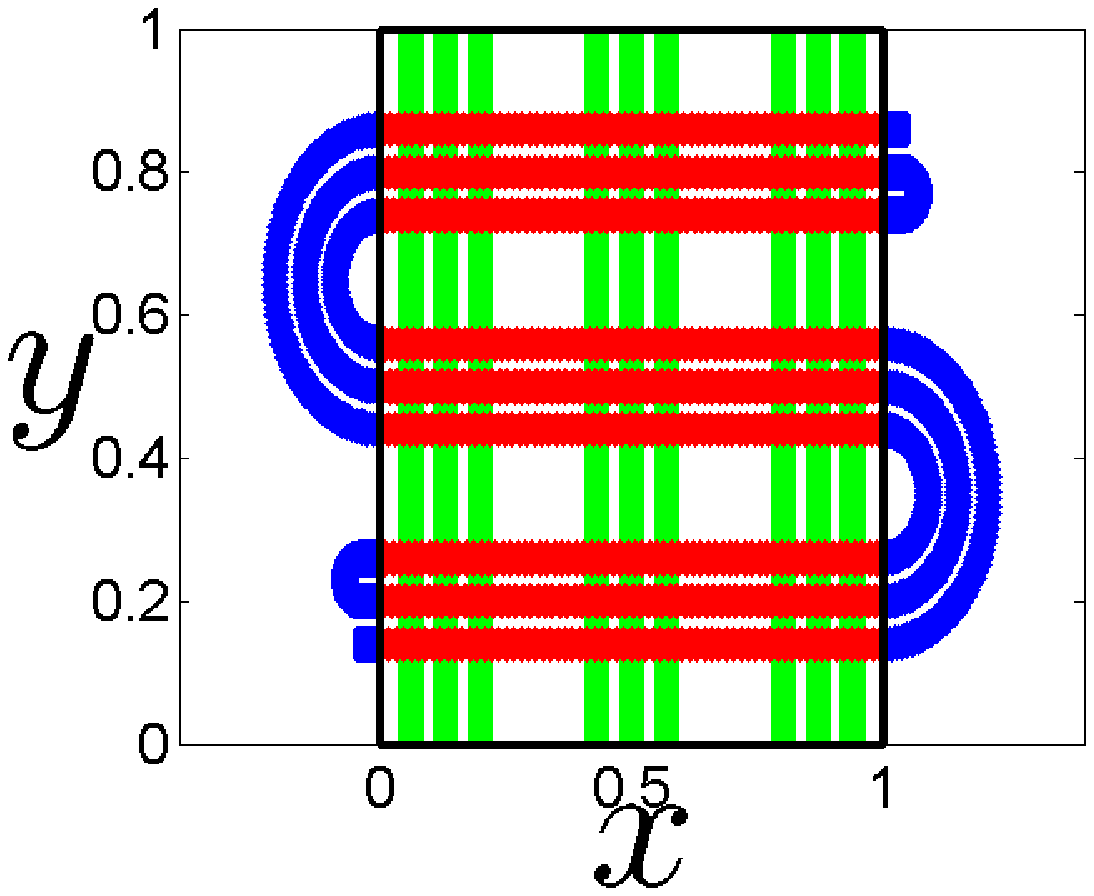}
   \label{subfig:h22}
   } 

\end{tabular}
\caption{(a) The first iteration of the unit square under the two-fold horseshoe map $G(Q)$ is shown in red and blue. The preimages of the red rectangles are shown in green. (b) The second iteration under the same map $G^{2}(Q)$.}
\label{fig:smho2}
\end{figure}

We represent the iterations of the unit square under the map $G=G_{4}\circ G_{3}\circ G_{2}\circ G_{1}$ in Fig.~\ref{fig:smho2}. Two new elementary transformations appear in $G_{3}$, which has an additional clockwise rotation for the new blue folding and also one more translation for the upper red rectangle appearing in Fig.~\ref{fig:smho2}(a). Now there are two stable fixed points out of the unit square, one for positive values of $x$ in the upper part, and another for negative values of $x$, in the lower part. As we show in the next section, the stable manifold of a saddle fixed point in the unit square separates the basin boundaries of these two attractors, whose boundary is fractal. 

We note that, for a horseshoe with a single folding, the number of attractors outside $Q$ is one, while in this case there are two. In fact, the same holds if the number of foldings is even, while there is only one attractor if the number of foldings is odd. Therefore, the coexistence of more than two attractors requires more than one folding horseshoe.

\section{Applications} \label{sec:pro}

Since chaos is related to the existence of horseshoes, the practical use of these piecewise maps is to investigate properties of chaotic dynamical systems. In the following lines we introduce several applications of these horseshoe maps, to illustrate how they can be used with different purposes. After examining two well-known examples, and as a completely new feature, we show that a three-fold horseshoe can give rise to basin boundaries exhibiting the so-called  Wada property \cite{yoney}. Given the fact that homoclinic tangles give rise to infinitely many successive foldings, this fact suggests why the Wada property appears so frequently in nonlinear multistable dynamical systems and dispersive systems with several escapes \cite{aguirre,vander,toka, wadye, wadel,chasca}. Finally, we show how one-dimensional unimodal or multimodal maps can be in general computed from horseshoes with one or more foldings, which can simplify enormously the manipulation of a dynamical system.

\subsection{Basins of attraction} \label{sec:bas}

\begin{figure}[ht]
\centering
\begin{tabular}{cc}

\subfigure[]{
   \includegraphics[width=0.372\linewidth]{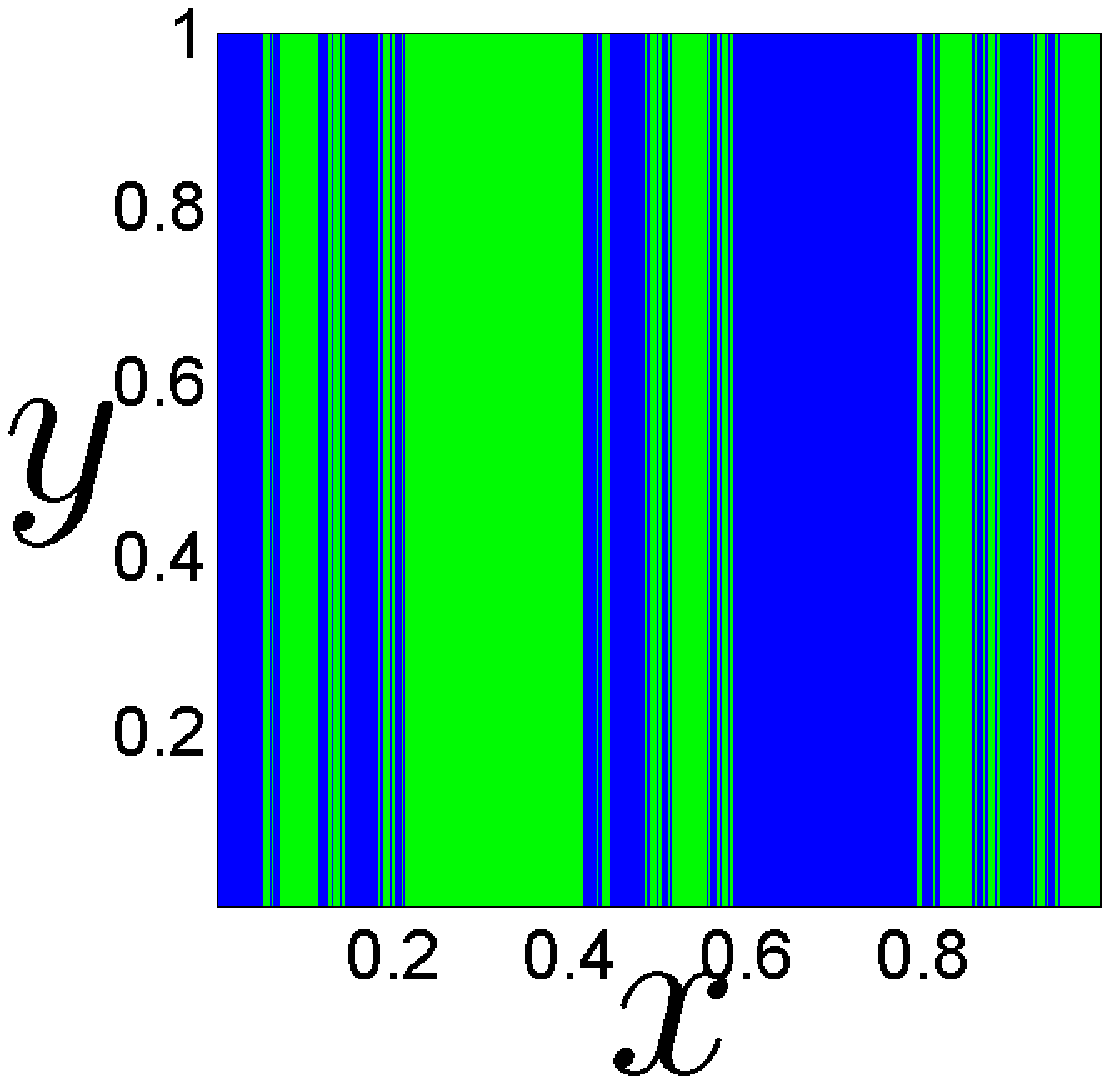}
   \label{subfig:bas1}
   } &
\subfigure[]{
   \includegraphics[width=0.36\linewidth]{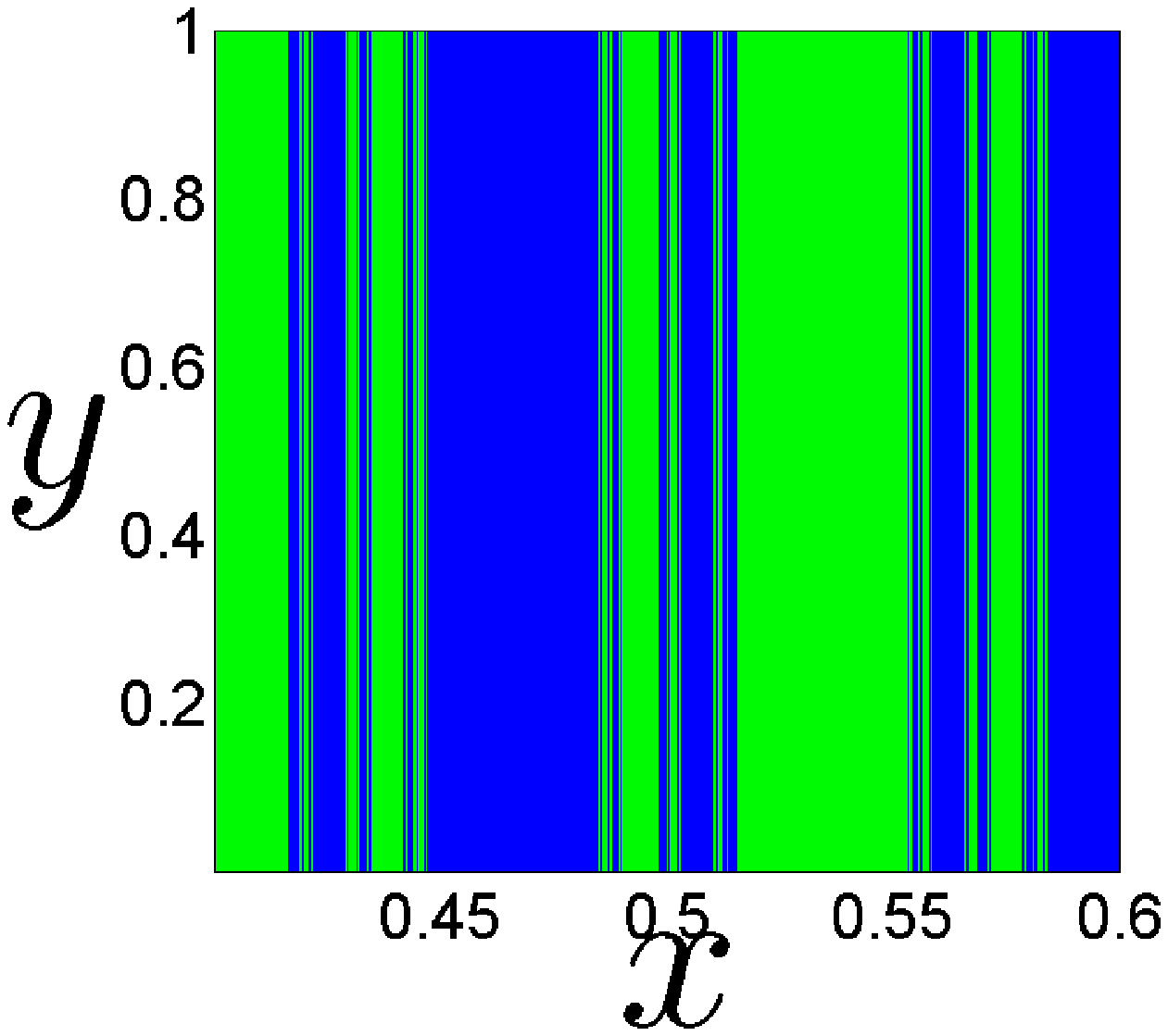}
   \label{subfig:bas2}
   } \\
\subfigure[]{
   \includegraphics[width=0.39\linewidth]{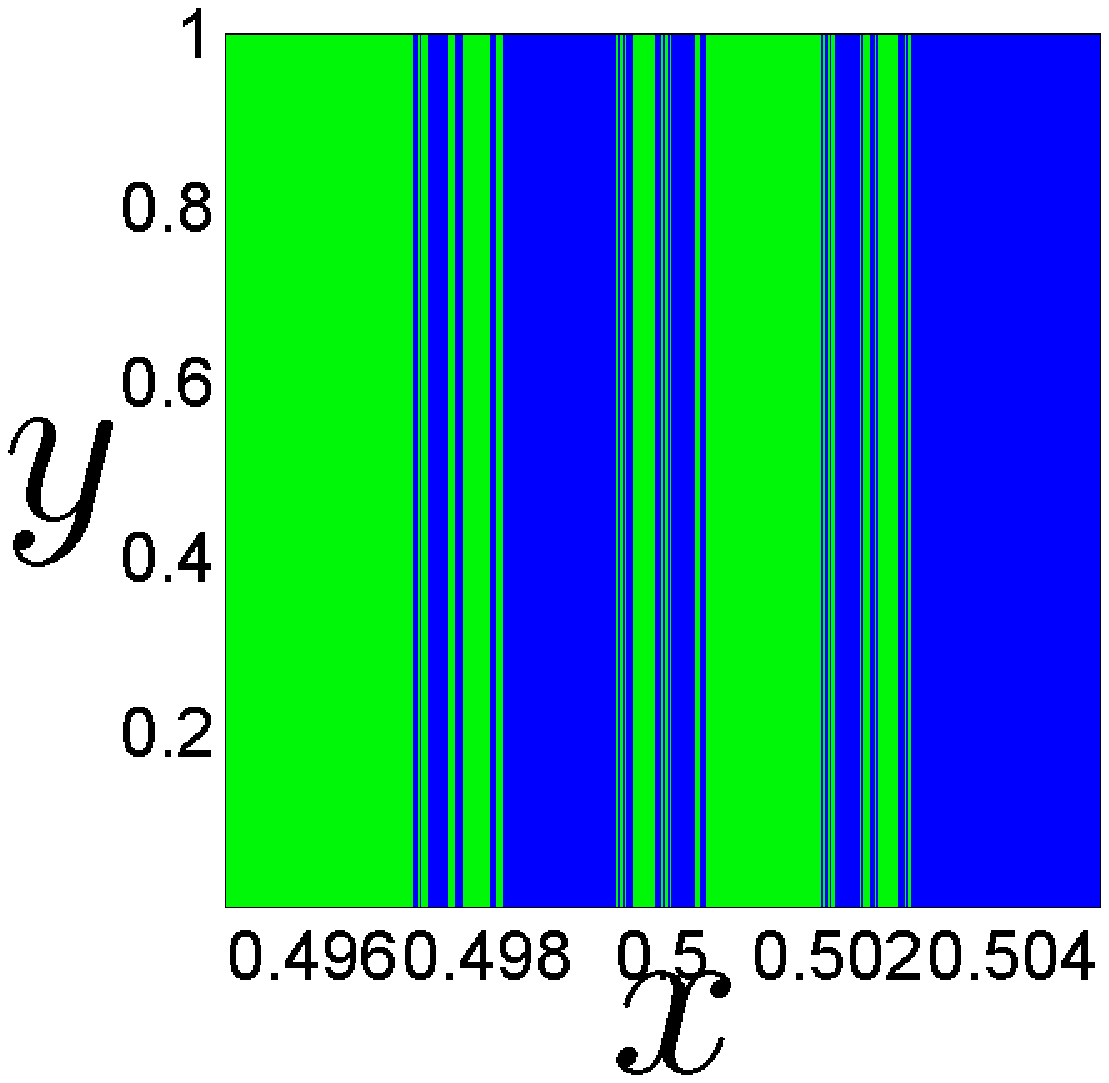}
   \label{subfig:bas3}
   } &
\subfigure[]{
   \includegraphics[width=0.365\linewidth]{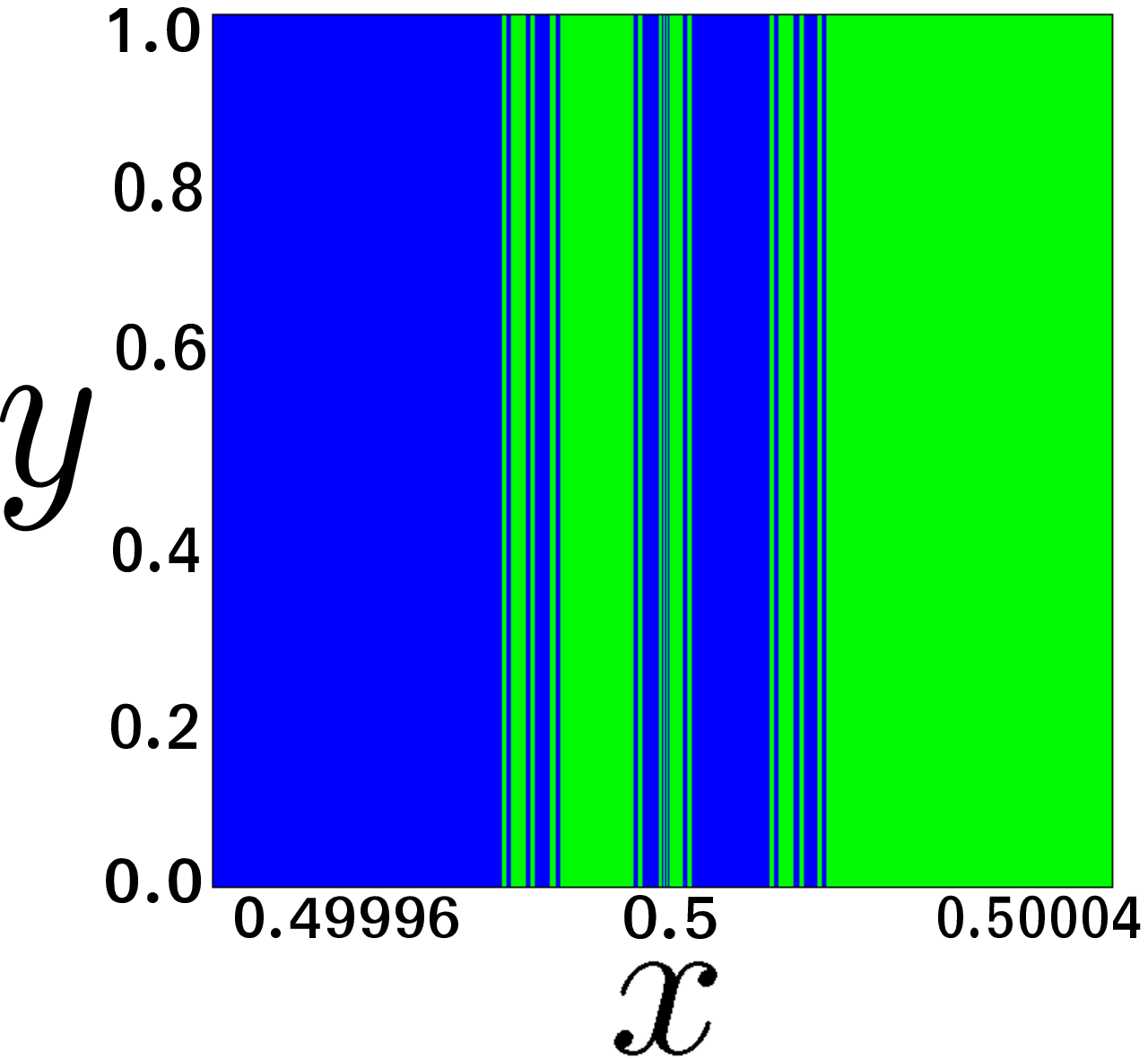}
   \label{subfig:bas4}
   }

\end{tabular}
\caption{(a) Basins of attraction of the two attractors $x^{*}_{1}=(-0.05,0.125)$ and $x^{*}_{2}=(1.05,0.875)$ of the two-fold horseshoe map $G$. (b)-(d) Three blow-ups of the basins showing their fractality. Note the symmetry of the basins on both sides of $x=1/2$, which is due to the symmetry of the stable manifold.}
\label{fig:bas}
\end{figure}

As shown in the previous section, the scheme proposed in \cite{macdonal} to demonstrate that horseshoes are responsible for the fractal nature of the boundary of the basins of attraction separating coexisting attractors, can be directly tested using these maps. We recall that a basin of attraction is the set of all the points that end in a particular attractor. In Fig.~\ref{fig:bas}(a) we represent the basins of attraction of the two attractors $x^{*}_{1}=(-0.05,0.125)$ and $x^{*}_{2}=(1.05,0.875)$ that lie outside the unit square given by the map $G$ described in the previous section. Their boundary consists of vertical lines, associated to the stable manifold of a saddle fixed point in the unit square $Q$. Successive magnifications of the basins are shown in Figs.~\ref{fig:bas}(b-d), which clearly suggest their fractal nature.

\subsection{Chaotic saddle} \label{sec:cas}

\begin{figure}[ht]
\centering
\bigskip
\bigskip
   \includegraphics[width=0.46\linewidth] {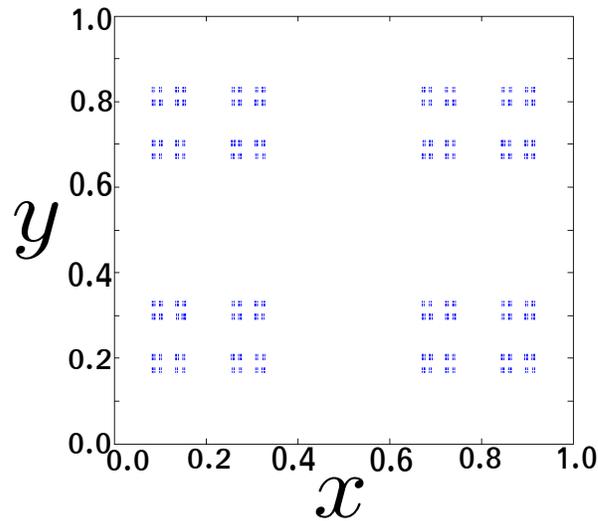} 

\caption{An approximation of the chaotic saddle of $H$ using the Sprinkler Algorithm with $n_{0}=8$, starting with a planar grid of resolution $60000 \times 60000$. This invariant set consists of the direct product of two Cantor sets.}
\label{fig:chasa}
\end{figure}

Another important invariant set associated to transient chaotic dynamics is the non-attracting invariant set, also called the chaotic saddle. This set is formed by all the points in the unit square that do not escape to the attractor by forward or backward iteration. It can be computed by means of the Sprinkler Algorithm \cite{sprinkal}, as shown in Fig.~\ref{fig:chasa}. If we look at Fig.~\ref{subfig:h4}, the chaotic saddle can be iteratively computed as the intersection $H^{n}(Q) \cap H^{-n}(H^{n}(Q)\cap Q)$ when $n$ tends to infinity and with $Q$ the unit square. For instance, the first iteration is the intersection of the green vertical and red horizontal strips in Fig.~\ref{subfig:h1}, which consists of four squares. 

\subsection{Basins of Wada} \label{sec:wabas}

Even though it is not possible to have more than two attractors outside the unit square with a horseshoe map, we can create systems with more than two escapes from the unit square and study the escape basins. In fact, using a minimum number of three foldings, it is possible to create an escape basin with the Wada property. Three basins possess this property if every point of the boundary of a particular basin belongs to the boundary of the remaining two basins \cite{ken,nus}.

\begin{figure}
\centering
\begin{tabular}{cc}

\subfigure[]{
   \includegraphics[width=0.5\linewidth] {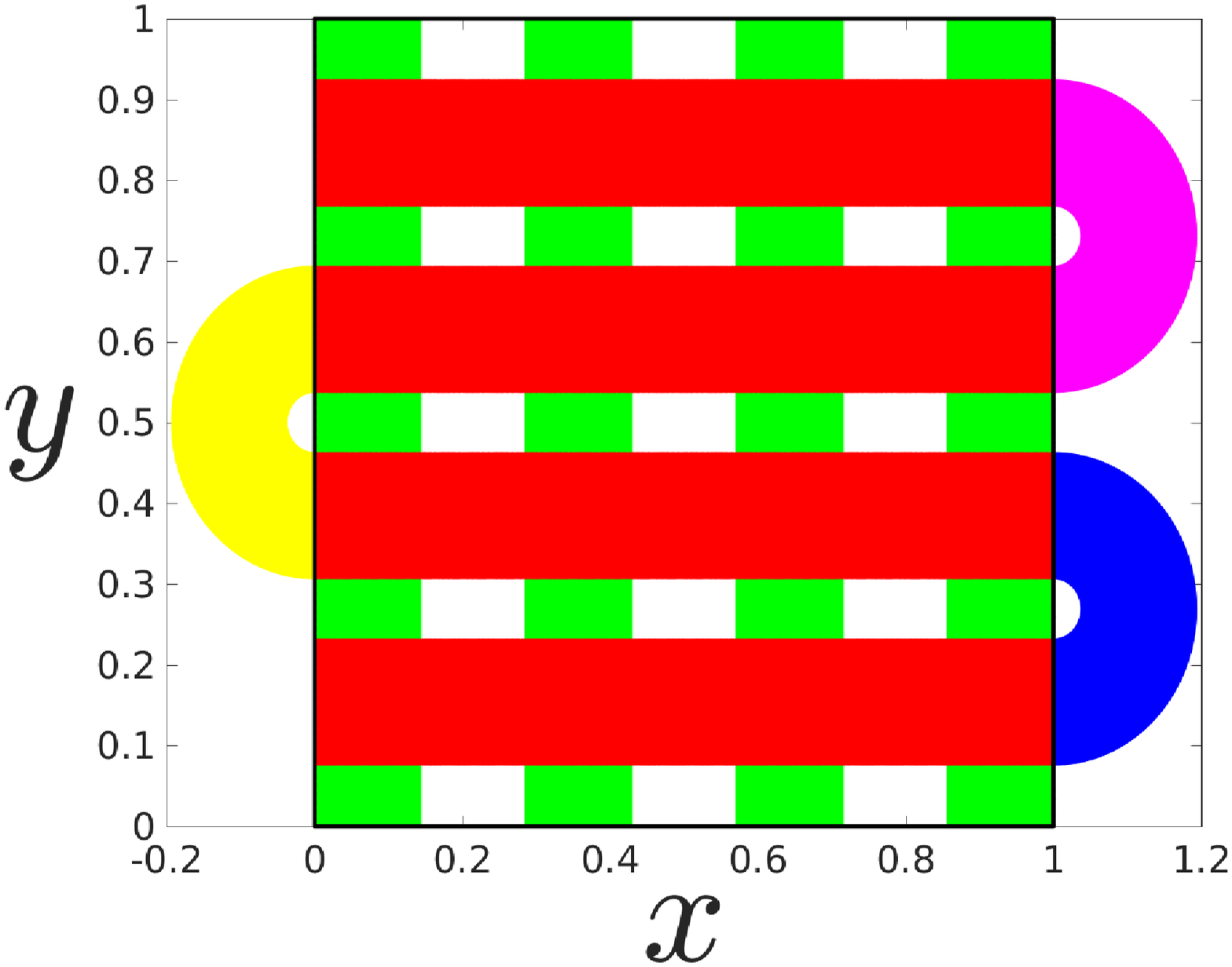}
   \label{subfig:h31}
   } &
\subfigure[]{
   \includegraphics[width=0.42\linewidth] {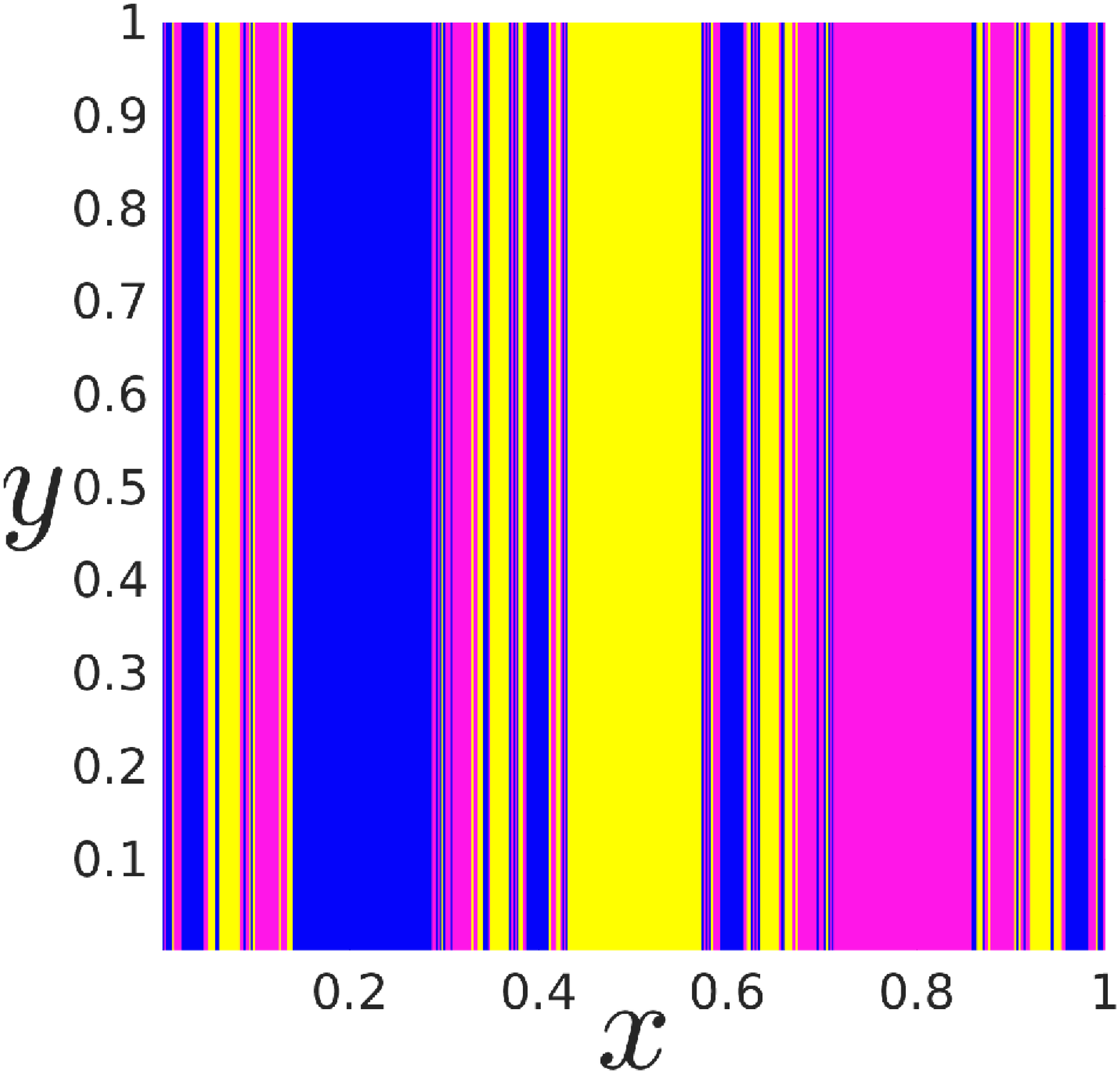}
   \label{subfig:h32}
   } 

\end{tabular}
\caption{(a) The first iteration of the unit square by means of a three-fold horseshoe is shown in red, blue, yellow and magenta. The preimages of the red rectangles are shown in green. (b) The escape basins, considering as different escapes each of the three foldings, with their corresponding colors. These escape basins possess the Wada property.}
\label{fig:smho3}
\end{figure}

A three-fold horseshoe map is shown in Fig.~\ref{fig:smho3}(a). The equations that describe this map are a straightforward extension of the equations shown in the previous section and we omit them because they are quite long. The escape basins can be defined, relating a different escape to each of the three foldings appearing in Fig.~\ref{fig:smho3}(a). The associated basin appearing in Fig.~\ref{fig:smho3}(b) is a basin of Wada, as can be demonstrated using recent numerical techniques \cite{test,merge}.

\subsection{Multimodal 1D maps} \label{sec:hormod}

\begin{figure}[H]
\centering
   \includegraphics[width=0.95\linewidth] {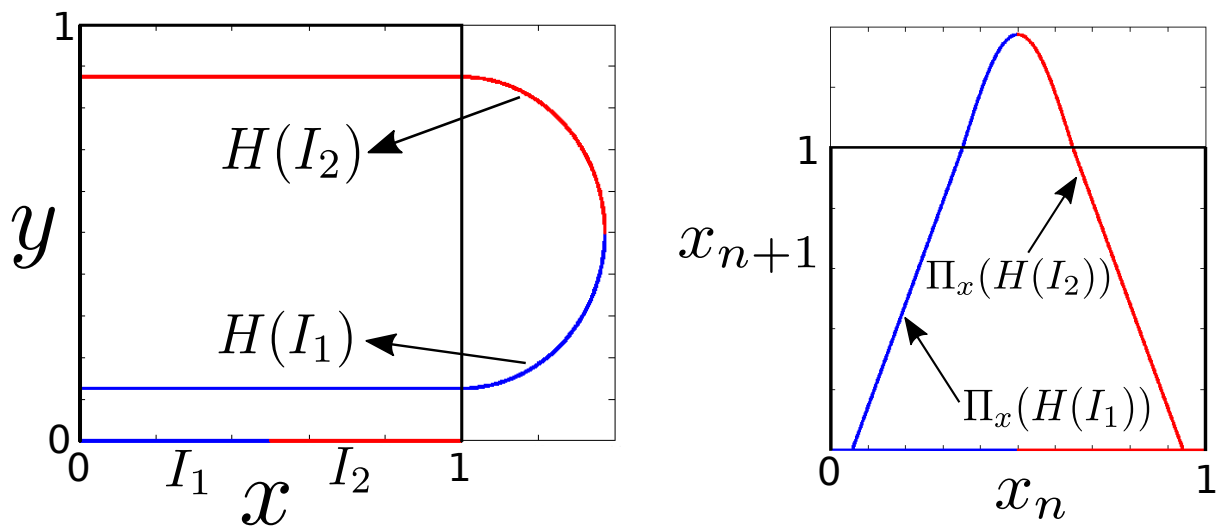} 
\caption{(a) The iteration of the lower side of the unit square by means of the piecewise horseshoe map. (b) The projected map on the $x$-axis, which yields a unimodal map, very similar to the logistic map.}
\label{fig:proyec}
\end{figure}

We conclude the present work by demonstrating that a horseshoe with multiple foldings yields a multimodal map, by simply projecting its boundary along the stretching direction. The relevance of this correspondence is that it allows to reduce the dimensionality of a dynamical system, which simplifies enormously its study. To the best of our knowledge, this fact appears sometimes in the literature, but without a detailed explanation of how these two maps are connected \cite{nicols}. As shown in Fig.~\ref{fig:proyec}(a), if we care about the lower edge $I=[0,1]$ of the unit square $Q$ only, we see that after one iteration $H(I)$ this set corresponds to the external boundary of the horseshoe. If we now project such external boundary on the $x$-axis by means of a projector $\Pi_{x}$, we obtain a unimodal map. In other words, the restriction of the horseshoe to the stretching axis (the $x$-axis in our examples) yields a unimodal one-dimensional map. As shown in Fig.~\ref{fig:proyec}(b), this fact is immediately verified using our piecewise maps. The present argument can be extended in a very simple way to show that a horseshoe with several foldings can be related to a multimodal map.

\section{Conclusions and Discussion} \label{sec:discussions}

We have provided a systematic procedure to compute complex folding structures. This is achieved by using two-dimensional piecewise maps, which accomplish all the required elementary transformations, such as translations, projections, scalings and rotations, both linear and nonlinear. These mathematical functions can be useful to investigate central properties of chaotic dynamical systems and, from the point of view of computational running times, they are very affordable.

A particularly interesting application has been presented, which consists, perhaps, in the conceptually simplest two-dimensional dynamical system that presents the Wada property. As we have shown, more than two escape basins can share their whole boundary if horseshoes with more than two foldings are present in the system. In particular, a folding structure with three foldings is sufficient for a dynamical system to display such property.

Another application of these maps is the illustration of how multimodal maps can be derived from horseshoe maps with several foldings. Despite its simplicity, this equivalence is of paramount importance. Methods to obtain a Smale horseshoe in an arbitrary low-dimensional chaotic dynamical system have been developed recently \cite{findho,findho2}. Therefore, this correspondence allows us to reduce the dimensionality of the dynamical system even further, and also to apply our vast knowledge on unidimensional maps \cite{may,shark,collet} to a wide spectrum of dynamical systems. As it has been recently shown, this capacity to reduce the dimensionality of the dynamical system can also entail a considerable simplification of some control methods, as for example the partial control method \cite{capeans}.

\section{Acknowledgements}  \label{sec:aknow}

This work has been supported by the Spanish State Research Agency (AEI) and the European Regional Development Fund (FEDER) under Project No. FIS2016-76883-P.

\end{document}